\documentclass[twocolumn,pra,aps]{revtex4}
\usepackage{physics}
\usepackage{graphicx,tikz}
\usepackage{algorithm2e}
\usepackage{amsmath,amssymb,amsfonts,amsthm}
\usepackage[colorlinks=true, citecolor=blue, urlcolor=blue ]{hyperref}
\usepackage{multirow}
\usepackage{verbatim}
\usepackage{url}
\usepackage{comment}
\usepackage{mathtools}
\usepackage{color}

\begin{document}
\title{Quantum Signaling to the Past Using P-CTCS\footnote{Part of this work was initiated when the first two authors visited R. C. Bose Centre for Cryptology and Security, Indian Statistical Institute, Kolkata during the Summer of 2017 (between the 6th and the 7th semester of their Bachelor of Engineering) for internship under the supervision of the third author.}}
\author{Soumik Ghosh$^1$, Arnab Adhikary$^1$ and Goutam Paul$^2$}
\affiliation{$^1$Department of Electronics \& Telecommunication Engineering,\\
Jadavpur University, Kolkata 700032, India.\\
Email: \{ghoshsoumik14, arnab45264\}@gmail.com\\
$^2$Cryptology and Security Research Unit,\\
R. C. Bose Centre for Cryptology and Security,\\
Indian Statistical Institute, Kolkata 700108, India.\\
Email: goutam.paul@isical.ac.in
}

\begin{abstract}
Closed Timelike Curves (CTCs) are intriguing relativistic objects that allow for time travel to the past and can be used as computational resources. In Deutschian Closed Timelike Curves (D-CTCs), due to the monogamy of entanglement, non-local correlations between entangled states are destroyed. In contrast, for Postselected Closed Timelike Curves (P-CTCs), a second variant of CTCs, the non-local correlations are preserved. P-CTCs can be harnessed for the signaling of non-orthogonal states to the past without a disruption of causality. In this paper, we take up signaling to the past and show a method of sending four non-orthogonal states to the past using P-CTCs. After constructing our signaling protocol, we study the causality violations that our protocol results in and put forward two consistency relations to prevent them.  
\end{abstract}

\keywords{Closed Timelike Curves, Quantum Entanglement, Superluminal Signaling}

\maketitle{}
\date{}

\section{Introduction}
Closed Timelike Curves arise out of Einstein's field equations of General Theory of Relativity \cite{Godel} and permit voyages to the past. Notwithstanding the deluge of science fiction concerning time travel to the past prevailing in popular culture, existence of CTCs remains a source of persistent debate in the scientific community. Putting that debate aside, David Deutsch, in his seminal paper, \cite{Deutsch} tried to model such a Closed Timelike Curve using principles of quantum interactions and information theory. The CTCs which abide by Deutsch's prescription are called Deutschian-CTCs. \par 
Brun, Harrington and Wilde~\cite{BHW} showed that non-orthogonal states can be perfectly distinguished using a D-CTC. This fact made quantum key distribution protocols, like BB84 \cite{Bennett}, untenable in a world with CTCs because they relied on non-distinguishability of arbitrary non-orthogonal states as their security measure.

As far as CTCs are concerned, Bennett and Schumacher \cite{Schumacher} suggested an alternative nonequivalent formulation of CTCs using quantum teleportation and post-selection (P-CTCs). Svetlichny \cite{Svetlichny} developed on this framework to suggest his own version of a time travelling circuit using teleportation. This was further developed by Lloyd et al.~\cite{Lloyd1} and experimentally simulated \cite{Lloyd} by Aephraim Steinberg's group at Toronto. Open Timelike Curves, a special case of D-CTCs, were also studied by \cite{OTCfirst, Yuan}.

Brun et al.~showed \cite{BrunPCTC} that a different variant of the Brun-Harrington-Wilde (BHW) circuit \cite{BHW} can distinguish non-orthogonal quantum states for P-CTCs, with only added constraint: such a circuit could only distinguish non-orthogonal states which were linearly independent. Hence, P-CTCs could no longer distinguish between the four BB84 states. In the meantime, Ralph proposed \cite{Ralph} a circuit for P-CTCs which can effectively signal to the past. Bub et al.~\cite{Bub} studied Ralph's circuit for causality and put forward two consistency conditions to preserve causality in such a circuit. Utilizing elements of quantum field theory and relativity, a quantum key distribution scheme was proposed by Ralph et al.~\cite{Ralphkey}. Recently, it was shown by Moulick and Panigrahi \cite{Moulick} that CTCs can increase entanglement using local operations and classical communications. 

In this paper, we first briefly discuss exchange of BB84 states using D-CTCs. Then, we borrow Ralph's circuit for P-CTCs, modify and correct Bub's \cite{Bub} proposal to transmit four non-orthogonal states using it, and finally discuss causality flow in such a circuit. Finally, we state two consistency conditions as sanity check, to preserve the flow of causality and prevent paradoxes.

Since CTCs are not much more than a theoretical construct, practical implementation of such a circuit is not feasible. Nonetheless, it is a good thought experiment which throws up some interesting questions regarding how causality plays out for someone tinkering with a CTC. We have highlighted and addressed those questions in this paper.

\section{Previous works and motivation}

In both variants of CTCs, quantum entanglement plays a crucial role in designing the signaling protocols \cite{Bub}. 

However, time travel comes with its assorted paradoxes and CTCs are not immune to the same. While designing the signaling protocols, we should ensure causality is not violated, giving rise to paradoxical results.

\subsection{Signaling using D-CTCs}
\begin{figure}[h]
\centering
\includegraphics[width=0.5\textwidth]{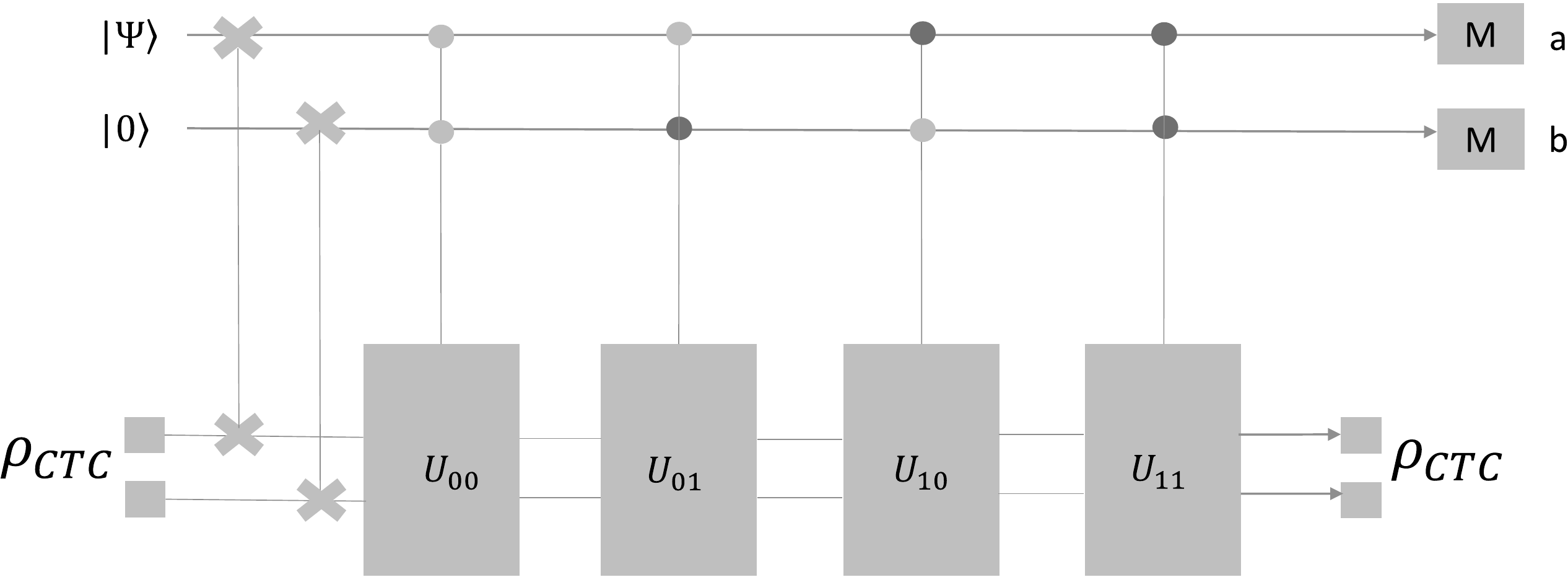}
\caption{The BHW circuit~\cite{BHW}. It is used for distinguishing the four BB84 states in the case of D-CTCs.}
\label{BB84}
\end{figure}

Using D-CTCs for quantum key distribution is trivial, as it immediately follows from the results of \cite{BHW} and \cite{Bub}. A BHW circuit~\cite{BHW} at Bob's end, as shown in Fig.~\ref{BB84}, is enough to detect what bit Alice is sending him from the future. The causality is also not violated, as discussed in \cite{Bub}, if we apply the two consistency conditions the authors proposed.

The challenge lies in extending the idea to Postselected Closed Timelike Curves as well. The two variants of CTCs are not alike and interact differently with entangled states. While D-CTCs decorrelate quantum states passing through it \cite{Deutsch}, P-CTCs preserve the entangled nature of the states \cite{Lloyd}. Hence, a causally consistent signaling protocol in one does not guarantee a causally consistent signaling protocol in another.

\subsection{Signaling using P-CTCs}
\begin{figure}[h]
\centering
\includegraphics[width=0.5\textwidth]{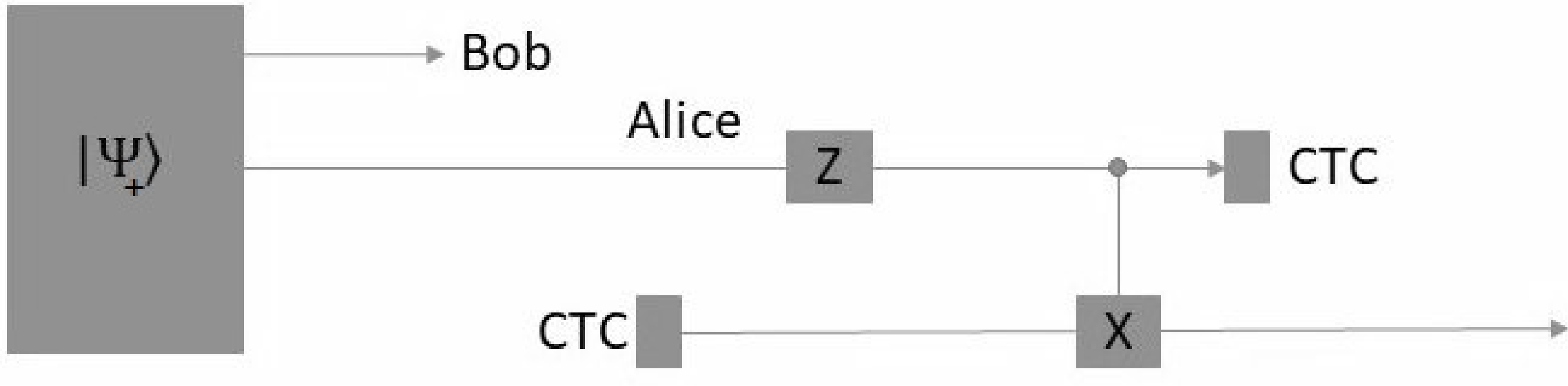}
\caption{Signaling to the past using P-CTCs~\cite{Ralph}. $\ket{\psi_+}$ is the entangled pair shared between Alice and Bob. Alice has two gates at her disposal, a phase flip gate and a CNOT gate. She passes her share of the entangled state through a postselected closed timelike curve. The state comes out from the other mouth of the P-CTC to become the target qubit of the controlled NOT operation. If Alice does not apply the phase flip gate, it can be shown that $\ket{+}$ gets transmitted to Bob in the past. If she applies the phase flip gate, $\ket{-}$ gets transmitted.}
\label{Ralph}
\end{figure}

\begin{figure}[h]
\centering
\includegraphics[width=0.5\textwidth]{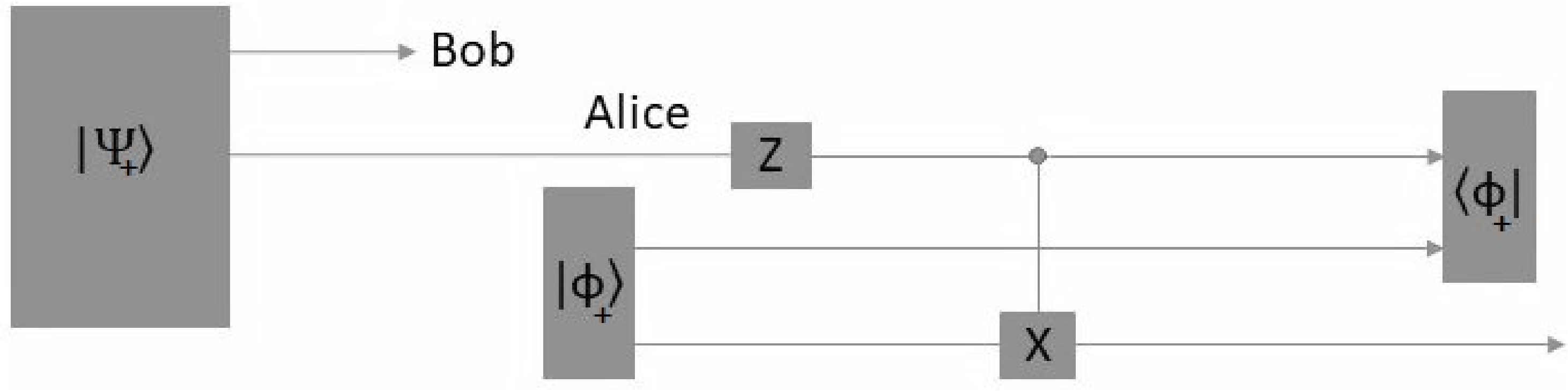}
\caption{The equivalent circuit of the ``radio to the past". $\ket{\phi_+}$ represents the Bell state imitating the P-CTC operation as per \cite{Lloyd}. In other words, the P-CTC action has been modeled using another pair of entangled qubits $\ket{\phi_+}$. Alice's share of $\ket{\psi_+}$ and one qubit of $\ket{\phi_+}$ are postselected to $\ket{\phi_+}$. The other qubit of $\ket{\phi_+}$ forms the target qubit of the controlled NOT operation.}
\label{Ralph2}
\centering
\end{figure}

A circuit for superluminal signaling was proposed by Ralph \cite{Ralph} in Fig.~\ref{Ralph}. The causality analysis of the circuit was done by Bub and Stairs~\cite{Bub}. The circuit is given in Fig.~\ref{Ralph}. The equivalent circuit using Lloyd's prescription is given in Fig.~\ref{Ralph2}.

Bub et al.'s calculations on P-CTCs are somewhat different from ours. In their analysis of causality, they applied the CNOT gate with Alice's qubit as control and the first half of the entangled Bell state $\ket{\phi_+}$ as target. As shown in Fig.~\ref{Ralph2}, in Ralph's equivalent circuit, the gate is applied with Alice's qubit as the control and the second half of the entangled pair $\ket{\phi_+}$ as the target. We have followed Ralph's convention in our analysis. Ralph had provided a circuit theoretic framework for such a transfer of a pair of states, which we extend by adding another gate for the transfer of four states.

We suitably modify the circuit Ralph has shown to transmit four non-orthogonal states to the past. Ralph had provided a circuit theoretic framework for such a transfer of a pair of states, which we extend by adding another gate for the transfer of four states. Even though Brun and Wilde in~\cite{BrunPCTC} had indicated in the mechanism for such a transfer, no extensive analysis has yet been done to probe the causality relations during the transfer of four non-orthogonal states to the past through a P-CTC. 

Our results are a natural extension of the results of \cite{Bub} and \cite{Ralph}. Bub et al. had formulated two causality relations during state transfer using D-CTCs. They had remarked that P-CTCs are self-consistent; hence the extraneous consistency relations would be redundant in their case. We show here that while P-CTCs are indeed self-consistent, additional consistency conditions are required for P-CTCs too, as sanity checks to avoid causal violations.

\section{Our Results}
\begin{figure}[h]
\centering
\includegraphics[width=0.5\textwidth]{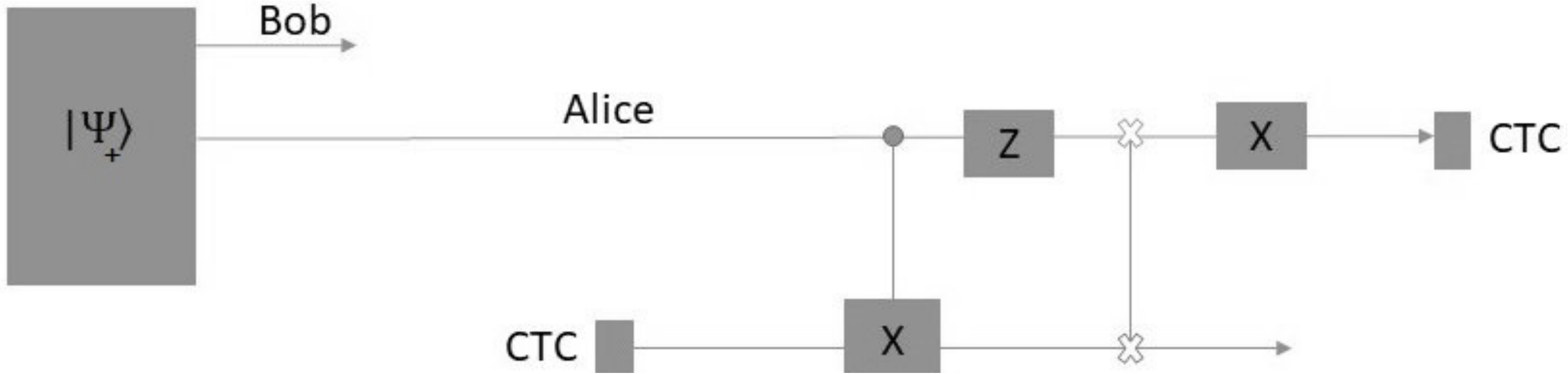}
\caption{The schematic of our proposal. Apart from the phase flip and controlled-NOT gates, we have used a SWAP gate and a NOT gate in order to transmit four non orthogonal states to the past, namely $\ket{0}$, $\ket{1}$, $\ket{+}$, $\ket{-}$.}
\label{Fig4}
\end{figure}

\begin{figure}[h]
\centering
\includegraphics[width=0.5\textwidth]{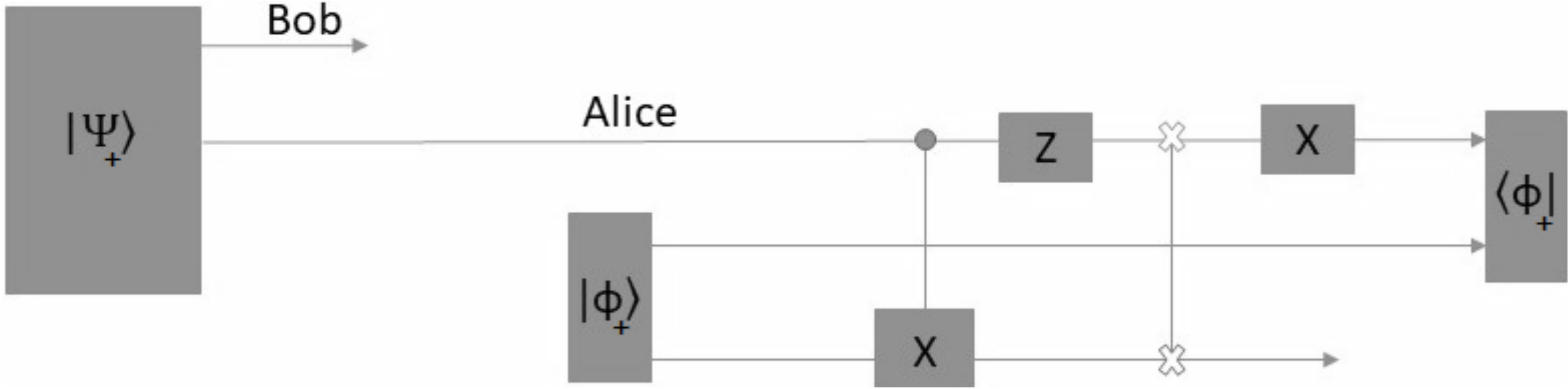}
\caption{The equivalent circuit of our proposal.}
\label{Fig5}
\end{figure}

Let Alice and Bob share a Bell state initially, each given by $\frac{1}{\sqrt{2}}(\ket{0_{B}}\ket{1_{A}} + \ket{1_{B}}\ket{0_{A}})$. 
Ralph et al. proposed a circuit \cite{Ralph} by which Alice can send the states $\ket{+}$ or $\ket{-}$  to the past using Postselected Closed Timelike Curves. We modify Ralph's circuit and use it to transfer all states of the BB84 protocol to the past. Our circuit is given in Fig.~\ref{Fig4}. The equivalent model of the circuit of Fig.~\ref{Fig4} following Lloyd's prescription \cite{Lloyd} is given in Fig.~\ref{Fig5}.

\subsection{Correctness of our signaling protocol}
Alice has at her disposal three components that she can freely apply. There is a phase flip gate, a controlled not gate, and a swap gate. Using one, or some of these components she is successfully able to signal to Bob in the past.
\begin{itemize}
\item Let us assume that Alice wants to send $\ket{1}$. She then applies the controlled NOT gate and the SWAP gate.

The initial state was \begin{equation*}\ket{\psi_+} = \frac{1}{\sqrt{2}}(\ket{0_{B}}\ket{1_{A}} + \ket{1_{B}}\ket{0_{A}}).\end{equation*}

The P-CTC interaction can be modeled by \begin{equation*}\ket{\phi_+} = \frac{1}{\sqrt{2}}(\ket{0_{1}}\ket{0_{2}} + \ket{1_{1}}\ket{1_{2}}).\end{equation*} Here, $1$ and $2$ denote the first and second halves respectively of the entangled state $\ket{\phi_+}$ in Fig.~\ref{Fig5}.

The total state initially is
\begin{eqnarray*}
\ket{\psi_+} \otimes \ket{\phi_+}
& = & \frac{1}{\sqrt{2}}(\ket{0_{B}}\ket{1_{A}} + \ket{1_{B}}\ket{0_{A}})\\
&& \otimes \frac{1}{\sqrt{2}}(\ket{0_{1}}\ket{0_{2}} + \ket{1_{1}}\ket{1_{2}}).
\end{eqnarray*}

After the CNOT gate, the state is 
\begin{eqnarray*}
\ket{\psi_+} \otimes \ket{\phi_+}
& = & \frac{1}{2}\ket{0_B}\ket{1_A} \otimes (\ket{0_1}\ket{1_2} + \ket{1_1}\ket{0_2})\\
&& 
+\frac{1}{2}\ket{1_B}\ket{0_A} \otimes (\ket{0_1}\ket{0_2} + \ket{1_1}\ket{1_2}).
\end{eqnarray*}

After the SWAP gate, the state is 
\begin{eqnarray*}
\ket{\psi_+} \otimes \ket{\phi_+}
& = & \frac{1}{2}\ket{0_B}\ket{1_A}\ket{0_1}\ket{1_2}\\
&& +  \frac{1}{2}\ket{0_B}\ket{0_A}\ket{1_1}\ket{1_2}\\
&& + \frac{1}{2}\ket{1_B}\ket{0_A}\ket{0_1}\ket{0_2}\\ 
&& +  \frac{1}{2}\ket{1_B}\ket{1_A}\ket{1_1}\ket{0_2}.
\end{eqnarray*}

Now, as per Fig.~\ref{Fig5}, she projects her share of the qubit and the first state of the P-CTC interaction to $\ket{\phi_+}$ and renormalizes. This leaves the total state as $\ket{1_B}\ket{1_2}$. Now, if Bob measures his state in the standard basis, he always gets 1.

\item Now, let us assume that Alice wants to send $\ket{0}$. This time, she applies a NOT gate after the SWAP gate. The total state, before the CNOT and the SWAP gates, is
\begin{eqnarray*}
\ket{\psi_+} \otimes \ket{\phi_+} & = & \frac{1}{2}\ket{0_B}\ket{1_A} \otimes (\ket{0_1}\ket{0_2} +  \ket{1_1}\ket{1_2}) \\
 & & + \frac{1}{2}\ket{1_B}\ket{0_A} \otimes (\ket{0_1}\ket{0_2} + \ket{1_1}\ket{1_2}).
\end{eqnarray*}

After the SWAP gate, the state is 
\begin{eqnarray*}
\ket{\psi_+} \otimes \ket{\phi_+}
& = &\frac{1}{2}\ket{0_B}\ket{1_A}\ket{0_1}\ket{1_2}\\
&&  +  \frac{1}{2}\ket{0_B}\ket{0_A}\ket{1_1}\ket{1_2}\\
&& + \frac{1}{2}\ket{1_B}\ket{0_A}\ket{0_1}\ket{0_2}\\
&& +  \frac{1}{2}\ket{1_B}\ket{1_A}\ket{1_1}\ket{0_2}.
\end{eqnarray*}

When she applies the NOT gate, the state is
\begin{eqnarray*}
\ket{\psi_+} \otimes \ket{\phi_+} 
& = & \frac{1}{2}\ket{0_B}\ket{0_A}\ket{0_1}\ket{1_2}  +  \frac{1}{2}\ket{0_B}\ket{1_A}\ket{1_1}\ket{1_2}\\
&&  + \frac{1}{2}\ket{1_B}\ket{1_A}\ket{0_1}\ket{0_2}  + \frac{1}{2}\ket{1_B}\ket{0_A}\ket{1_1}\ket{0_2}.
\end{eqnarray*}

Upon projection and renormalization, we are left with the final state $\ket{0_B}\ket{1_2}$, which is what Alice desired.

\item Alice now wants to send 0 in the diagonal basis, i.e., she wants to send $\ket{+}$. For this, she applies just the CNOT gate and renormalizes.

After the CNOT gate, the state is 
\begin{eqnarray*}
\ket{\psi_+} \otimes \ket{\phi_+} & = &\frac{1}{2}\ket{0_B}\ket{1_A} \otimes (\ket{0_1}\ket{1_2} + \ket{1_1}\ket{0_2})\\
&&+ \frac{1}{2}\ket{1_B}\ket{0_A} \otimes (\ket{0_1}\ket{0_2} + \ket{1_1}\ket{1_2}).
\end{eqnarray*}

Projection and renormalization gives $$\frac{1}{\sqrt{2}}(\ket{0_B} + \ket{1_B})\ket{0_2}.$$

\item To send 1 in the diagonal basis, she applies the phase flip after the CNOT. Now, the total state after application of these gates is
\begin{eqnarray*}
\ket{\psi_+} \otimes \ket{\phi_+} & = & -\frac{1}{2}\ket{0_B}\ket{1_A} \otimes (\ket{0_1}\ket{1_2} + \ket{1_1}\ket{0_2}) \\
& & + \frac{1}{2}\ket{1_B}\ket{0_A} \otimes (\ket{0_1}\ket{0_2} + \ket{1_1}\ket{1_2}).
\end{eqnarray*}

Projection and renormalization gives, neglecting the inconsequential global phase, \begin{equation*}\ket{\psi_+} \otimes \ket{\phi_+} = \frac{1}{\sqrt{2}}(\ket{0_B} - \ket{1_B})\ket{0_2}.\end{equation*}
\end{itemize}

\subsection{Analysis of causality: events at Bob's end}
Bob may not know which basis to measure his qubits in. Let us assume that Alice sent $1$ to Bob in the diagonal basis. But Bob used standard basis for his measurement. So, he gets $\ket{0_B}$ with probability $\frac{1}{{2}}$ and $\ket{1_B}$ with probability $\frac{1}{{2}}$. Let us assume that he gets $\ket{1_B}$. The initial total state now is 
\begin{equation*}\ket{\psi_+} \otimes \ket{\phi_+} =  \frac{1}{\sqrt{2}}(\ket{1_{B}}\ket{0_{A}}) \otimes \frac{1}{\sqrt{2}}(\ket{0_{1}}\ket{0_{2}} + \ket{1_{1}}\ket{1_{2}}).\end{equation*}

This state stays unchanged after the CNOT gate. As is evident from the state, the projection is $\ket{1_B}\ket{0_2}$, which means, the event that Alice and Bob use different basis for their measurement is not ruled out.

Now, let us assume Alice sent Bob $\ket{0}$ in the standard basis. But he used a diagonal basis for his measurement and got 0, ie $\ket{+}$. This would mean the initial state is \begin{equation*}\ket{\psi_+} \otimes \ket{\phi_+} = (\ket{+_{B}}\ket{+_{A}}) \otimes \frac{1}{\sqrt{2}}(\ket{0_{1}}\ket{0_{2}} + \ket{1_{1}}\ket{1_{2}}).\end{equation*}

The state after the CNOT and SWAP gate is 
\begin{equation*}
\begin{split}
\ket{\psi_+} \otimes \ket{\phi_+}  = \frac{1}{2}\ket{0_B}\ket{0_A}\ket{0_1}\ket{0_2}   +   \frac{1}{2}\ket{1_B}\ket{1_A}\ket{0_1}\ket{1_2}   \\ +  \frac{1}{2}\ket{0_B}\ket{1_A}\ket{0_1}\ket{1_2}    + 
\frac{1}{2}\ket{1_B}\ket{0_A}\ket{0_1}\ket{0_2}  \\ + 
\frac{1}{2}\ket{0_B}\ket{1_A}\ket{1_1}\ket{0_2}   +   \frac{1}{2}\ket{1_B}\ket{0_A}\ket{1_1}\ket{1_2} \\ + \frac{1}{2}\ket{0_B}\ket{0_A}\ket{1_1}\ket{1_2}  +  \frac{1}{2}\ket{1_B}\ket{1_A}\ket{1_1}\ket{0_2}.
\end{split}
\end{equation*}

After the NOT operation, the state is
\begin{equation*}
\begin{split}
\ket{\psi_+} \otimes \ket{\phi_+}  = \frac{1}{2}\ket{0_B}\ket{1_A}\ket{0_1}\ket{0_2}  +   \frac{1}{2}\ket{1_B}\ket{0_A}\ket{0_1}\ket{1_2} \\ + \frac{1}{2}\ket{0_B}\ket{0_A}\ket{0_1}\ket{1_2}  + 
\frac{1}{2}\ket{1_B}\ket{1_A}\ket{0_1}\ket{0_2} \\ + 
\frac{1}{2}\ket{0_B}\ket{0_A}\ket{1_1}\ket{0_2}  +  \frac{1}{2}\ket{1_B}\ket{1_A}\ket{1_1}\ket{1_2} \\ + \frac{1}{2}\ket{0_B}\ket{1_A}\ket{1_1}\ket{1_2}  +  \frac{1}{2}\ket{1_B}\ket{0_A}\ket{1_1}\ket{0_2}.
\end{split}
\end{equation*}

Upon projection and normalization, we get $\ket{+}\ket{1_2}$. These two cases are, quite obviously, an apparent contradiction, which throws up an interesting question about causality in P-CTCs. As Bub et al.~\cite{Bub} observed, P-CTCs require consistent events to occur due to the very way a P-CTC is defined. Here, the events which are occurring are indeed consistent, as mathematics shows. \par We can say that Bob's measurement in the wrong basis causes Alice to unknowingly send him $\ket{+}$ in the first place. After she prepares her state to send, Alice never measures her part of the qubit. She uses her ``radio to the past" with the inherent assumption that Bob will measure his qubit in the correct basis. When Bob does not do that, causality is, so to speak, reversed. Now, Bob measuring his qubit in the incorrect basis ``causes" Alice to send him a qubit which when measured in that basis produces no inconsistency and follows the rules of P-CTC. Even though Alice intends to signal in the basis of her choice, Bob's measurement forces the hand of nature and makes it such that the qubit sent is in the basis which does not contradict the principle of P-CTCs. \par

We can generalize the above result and show, considering each case, that whenever Alice decides to signal to Bob in a particular basis, and Bob, by mistake, measures in the wrong basis, there is a reversal of causality from normal operation.

\subsection{Analysis of causality: events at Alice's end}
The events at Alice's end need to be carefully observed just like events at Bob's end to ensure causality is not violated. Alice has free will and she can choose any gate she wishes for signaling, no matter what Bob measures. \par To illustrate, she can choose to apply the phase flip gate even when Bob has measured the state $\ket{+}$ with correct choice of basis. Or, she can choose not to apply the phase flip gate when Bob has correctly measured $\ket{-}$. Bub et al. showed that in both these cases, the projection subspace is null. As per standard interpretation of P-CTCs, this means such events won't happen. They invoke the ``banana peel" principle, meaning something or the other will happen, like Alice slipping over a banana peel \cite{Bub}, which will prevent her from doing something that is inconsistent. However, the projection subspace is no longer null in our case. \par

For example, let Bob get $\ket{1_B}$ and thereafter, Alice decides to apply just the CNOT gate and nothing else, meaning $\ket{1_B}$ cannot possibly be sent. This event, apparently, cannot occur as it would mean a disturbance in causality.

The state before projection is 
\begin{eqnarray*}
\ket{\psi_+} \otimes \ket{\phi_+}
& = & \frac{1}{\sqrt{2}}(\ket{1_{B}}\ket{0_{A}}) \\
& & \otimes \frac{1}{\sqrt{2}}(\ket{0_{1}}\ket{0_{2}} + \ket{1_{1}}\ket{1_{2}}).\end{eqnarray*}

As is evident, the projection subspace is $\ket{1_B}\ket{0_2}$.

Again, let us suppose that Bob got $\ket{1_B}$ but Alice applies the NOT gate after the CNOT gate and the SWAP gate.

The state before projection is
$$
\frac{1}{2}\ket{1_B}\ket{1_A}\ket{0_1}\ket{0_2} +  \frac{1}{2}\ket{1_B}\ket{0_A}\ket{1_1}\ket{0_2}.$$

The projection subspace is $\ket{1_B}\ket{0_2}$.

Now, let us suppose Bob got $\ket{+}$, but Alice applies the SWAP and NOT gates, meaning $\ket{+}$ cannot be sent.
The state before projection is
\begin{eqnarray*}
\ket{\psi_+} \otimes \ket{\phi_+}
& = &\frac{1}{2}\ket{+_B}\ket{1_A}\ket{0_1}\ket{0_2}  + \frac{1}{2}\ket{+_B}\ket{0_A}\ket{1_1}\ket{0_2} \\
&& + \frac{1}{2}\ket{+_B}\ket{0_A}\ket{0_1}\ket{1_2}  + \frac{1}{2}\ket{+_B}\ket{1_A}\ket{1_1}\ket{0_2}.
\end{eqnarray*}

The projection subspace is $\ket{+_B}\ket{+_2}$.

These are rather fascinating results which are not in agreement with the banana peel principle described earlier. Something should have stopped these events from occurring, but as is shown above, these events are perfectly consistent and can definitely occur. The only explanation here is again a reversal of causality. When Bob measures a state, which is not eliminated by post-selection, he ``causes" Alice to send that state to him in the past. \par But, quite queerly, the fact that Bob is measuring a state ``means" that Alice sent him that state from the future. If Alice had wanted to send Bob a different state, Bob would not have measured it in the first place. So, even though the solutions reversing causality are not mathematically ruled out, they seem contrary to the notion that Alice has free will at her end to send a state.
\subsection{Consistency relations}
We can invoke the two consistency conditions Bub et al. used for D-CTCs to clear out the redundant solutions. As we see here, they are valid and extremely essential for P-CTCs as well. \cite{Bub} The conditions are as follows. \par 
\begin{enumerate}
    \item Observers in differently moving reference frames must agree on which events occur, even if they disagree about the order of events. 
    \item  If an event has zero probability in any frame of reference, it does not occur. 
\end{enumerate}

Let's say for one observer time moves forward. He sees the events in the following order. 
\begin{enumerate}
\item Bob measures his state.
\item Alice applies operations on the disentangled state.
\end{enumerate}

The other observer is going backwards in time. For her:
\begin{enumerate}
\item Alice applies operations on the entangled state. 
\item Bob measures only after Alice has performed her unitaries.
\end{enumerate}

Here, to an observer who sees Alice applying her gates first, and Bob measuring his qubit after that, the probability that Alice applies the SWAP and NOT gates and Bob then measures $\ket{+}$ is 0. Hence, as it has 0 probability in this reference frame, it never occurs, by the second consistency condition, even though to an observer who sees Bob's actions first and then Alice's, it may seem consistent. The same principle holds true for other apparent inconsistencies.

\section{Conclusion}
We have developed and elaborated on a correct protocol for signaling to the past using Postselected Closed Timelike Curves (P-CTCs), and also remarked how it could be done using D-CTCs. We showed that it does not violate causality, and also provided the necessary mathematics for the same.  In all our results, quantum entanglement is a fundamental resource. \par Possible extensions to our work can look into ways of generalizing our circuit model to encode any arbitrary quantum state and do a causality analysis of the same. One can also look into applications of our scheme in probing different interesting features of P-CTCs.


\begin{thebibliography}{10}

\bibitem{Godel}
\bibinfo{author}{Godel, K.}
\newblock \bibinfo{journal}{\bibinfo{title}{An example of a new type of
  cosmological solutions of Einstein's field equations of gravitation.}}
{Rev. Mod. Phys.}
  \textbf{\bibinfo{volume}{21}}, \bibinfo{pages}{447--450}
  (\bibinfo{year}{1949}).

\bibitem{Deutsch}
\bibinfo{author}{Deutsch, D.}
\newblock \bibinfo{journal}{\bibinfo{title}{Quantum mechanics near closed
  timelike lines.}}
{Phys. Rev. D.} \textbf{\bibinfo{volume}{44}},
  \bibinfo{pages}{3197–3217} (\bibinfo{year}{1991}).

\bibitem{BHW}
\bibinfo{author}{Brun, T.~A.}, \bibinfo{author}{Harrington, J.} \&
  \bibinfo{author}{Wilde, M.~M.}
\newblock \bibinfo{journal}{\bibinfo{title}{Localized closed timelike curves
  can perfectly distinguish quantum states.}}
{Phys. Rev. Lett.}
  \textbf{\bibinfo{volume}{21}}, \bibinfo{pages}{210402}
  (\bibinfo{year}{2009}).

\bibitem{Bennett}
\bibinfo{author}{Bennett, C.} \& \bibinfo{author}{Brassard, G.}
\newblock \bibinfo{journal}{\bibinfo{title}{Quantum cryptography: Public key
  distribution and coin tossing.}}
{Proceedings of IEEE International Conference on
  Computers, Systems and Signal Processing} \bibinfo{pages}{175-179}
  (\bibinfo{year}{1984}).
  


\bibitem{Schumacher}
\bibinfo{author}{Bennett, C.}
Talk at QUPON, Wien. http://www.research.ibm.com/people/b/bennetc/
(\bibinfo{year}{2005}).

\bibitem{Svetlichny}
\bibinfo{author}{Svetlichny, G.}
\newblock \bibinfo{journal}{\bibinfo{title}{Effective Quantum Time Travel}}.
{arXiv:1204.4022v1}  (\bibinfo{year}{2009}).

\bibitem{Lloyd1}
\bibinfo{author}{Lloyd, S.}
\newblock \bibinfo{journal}{\bibinfo{title}{Quantum mechanics of time travel through post-selected teleportation.}}
{Phys. Rev. D}
\textbf{\bibinfo{volume}{84}} (\bibinfo{year}{2011}).

\bibitem{Lloyd}
\bibinfo{author}{Lloyd, S.}
\newblock \bibinfo{journal}{\bibinfo{title}{Closed timelike curves via post-selection: theory and experimental demonstration.}}
{Phys. Rev. Lett.}
\textbf{\bibinfo{volume}{106}} (\bibinfo{year}{2011}).

\bibitem{OTCfirst}
\bibinfo{author}{Pienaar, J.L., Myers, C.R., Ralph, T.C.}
\newblock \bibinfo{journal}{\bibinfo{title}{Open timelike curves violate Heisenberg's uncertainty principle}}.
{Physical Review Letters}
\textbf{\bibinfo{volume}{110}} (\bibinfo{year}{2013}).


\bibitem{Yuan}
\bibinfo{author}{Yuan, X. e.~a.}
\newblock \bibinfo{journal}{\bibinfo{title}{Replicating the benefits of Deutschian Closed Timelike Curves without breaking causality.}}
{Nature} \textbf{\bibinfo{volume}{15007}}
  (\bibinfo{year}{2015}).

\bibitem{BrunPCTC}
\bibinfo{author}{Brun, T.~A.} \& \bibinfo{author}{Wilde, M.~M.}
\newblock \bibinfo{journal}{\bibinfo{title}{Perfect state distinguishability
  and computational speedups with postselected closed timelike curves.}}
{Found. of Phys. Lett.}
  \textbf{\bibinfo{volume}{42}}, \bibinfo{pages}{341--361}
  (\bibinfo{year}{2012}).

\bibitem{Ralph}
\bibinfo{author}{Ralph, T.}
\newblock \bibinfo{journal}{\bibinfo{title}{Problems with modelling closed timelike curves as post-selected teleportation}}.
{arXiv:1107.4675}  (\bibinfo{year}{2011}).

\bibitem{Bub}
\bibinfo{author}{Bub, J.} \& \bibinfo{author}{Stairs, A.}
\newblock \bibinfo{journal}{\bibinfo{title}{Quantum interactions with closed
  timelike curves and superluminal signaling}}.
{Phys. Rev. A} \textbf{\bibinfo{volume}{89}}
  (\bibinfo{year}{2014}).
  
  
\bibitem{Ralphkey}
\bibinfo{author}{Ralph, T. and Walk, N.}
\newblock \bibinfo{journal}{\bibinfo{title}{Quantum key distribution without sending a quantum signal.}}
{New Journal of Physics}
\textbf{\bibinfo{volume}{17}} (\bibinfo{year}{2015}).


  \bibitem{Moulick}
\bibinfo{author}{Moulick, Subhayan} \& \bibinfo{author}{Panigrahi, Prasanta}
\newblock \bibinfo{journal}{\bibinfo{title}{Timelike curves can increase entanglement with LOCC.}}
{Scientific Reports} \textbf{\bibinfo{volume}{6}}
  (\bibinfo{year}{2016}).

  
  \bibitem{Wooters}
\bibinfo{author}{Wooters, W.K.} \& \bibinfo{author}{Zurek, W.H.}
\newblock \bibinfo{journal}{\bibinfo{title}{A single quantum cannot be cloned.}}
{Nature},
\bibinfo{pages}{802--803} (\bibinfo{year}{1982}).

  
  \bibitem{BennettNL}
\bibinfo{author}{Charles Bennett et al.}
\newblock \bibinfo{journal}{\bibinfo{title}{Can closed timelike curves or nonlinear quantum mechanics improve quantum state discrimination or help solve hard problems?}}
{Physical Review Letters} \textbf{\bibinfo{volume}{103:170502}}
  (\bibinfo{year}{2009}).
  
  \bibitem{Nagy}
\bibinfo{author}{Marius Nagy and Naya Nagy.}
\newblock \bibinfo{journal}{\bibinfo{title}{An Information-Theoretic Perspective on the Quantum Bit Commitment Impossibility Theorem.}}
{MDPI} (\bibinfo{year}{2018}).

 \bibitem{Cerf}
\bibinfo{author}{Nicolas J. Cerf and Chris Adami.}
\newblock \bibinfo{journal}{\bibinfo{title}{Negative entropy in quantum information theory.
}}
{New Developments on Fundamental Problems in Quantum Physics. } (\bibinfo{year}{1997})

  




\end{thebibliography}
\end{document}